**Observation of Anomalous Hall Effect in Noncollinear Antiferromagnetic Mn$_3$Sn Films**

*Yunfeng You, Xianzhe Chen, Xiaofeng Zhou, Youdi Gu, Ruiqi Zhang, Feng Pan and Cheng Song\**

Y. You, X. Chen, X. Zhou, Y. Gu, R. Zhang, Prof. F. Pan, Prof. C. Song

Key Laboratory of Advanced Materials (MOE), School of Materials Science and Engineering, Tsinghua University, Beijing 100084, People's Republic of China

*E-mail: songcheng@mail.tsinghua.edu.cn

Abstract: Magnetotransport is at the center of the spintronics. Mn$_3$Sn, an antiferromagnet that has a noncollinear 120° spin order, exhibits large anomalous Hall effect (AHE) at room temperature. But such a behavior has been remained elusive in Mn$_3$Sn films. Here we report the observation of robust AHE up to room temperature in quasi-epitaxial Mn$_3$Sn thin films, prepared by magnetron sputtering. The growth of both $(11\bar{2}0)$- and (0001)-oriented Mn$_3$Sn films provides a unique opportunity for comparing AHE in three different measurement configurations. When the magnetic field is swept along (0001) plane, such as the direction of $[01\bar{1}0]$ and $[2\bar{1}\bar{1}0]$, the films show comparatively higher anomalous Hall conductivity than its perpendicular counterpart ([0001]), irrespective of their respectively orthogonal current along [0001] or $[01\bar{1}0]$. A quite weak ferromagnetic moment of ~3 emu cm$^{-3}$ is obtained in $(11\bar{2}0)$-oriented Mn$_3$Sn films, guaranteeing the switching of the Hall signals with magnetization reversal. Our finding would advance the integration of Mn$_3$Sn in antiferromagnetic spintronics.

Antiferromagnets (AFM) show ultrafast spin dynamics with characteristic frequencies in the THz range, generate negligible stray fields and are robust against magnetic perturbations, offering prospects for the design of reliable high-density memories with fast operation speeds.[1–9] Robust magnetotransport properties are critical for the electronic device application of antiferromagnets.[2,3,10,11] The anomalous Hall effect (AHE), a typical magnetotransport behavior, is generally considered to have two possible origins: an extrinsic part attributed to energy-dissipative scattering as well as a dissipationless intrinsic part stemming from the Berry-phase curvature of the Bloch states in the momentum space, which is not surprising to vanish in collinear AFM.[12,13] However, in noncollinear chiral antiferromagnets, the nonvanishing Berry phase caused by the noncollinear spin arrangement has been theoretically and experimentally revealed, thus triggering a large AHE, reaching the same order of magnitude of ferromagnets even at zero magnetic field.[11,14–16] AHE of antiferromagnets connects magnetic and transport properties together, providing an effective way to explore the magnetic moments of antiferromagnets.

The hexagonal $Mn_3X$ antiferromagnets (where X = Ga, Ge, and Sn) are ideal systems to observe AHE due to the distinctive noncollinear 120-degree spin order,[11,17,18] among which, $Mn_3Sn$ is a kind of typical material firstly revealed to exhibit large AHE in single crystals experimentally.[11,14] As illustrated in Figure 1a, with the space group of P63/mmc, an inverse triangular spin structure is formed due to the geometrical frustration in the Kagome lattice of Mn atoms within the *a-b* plane and the Dzyaloshinskii-Moriya (DM) interactions below the Néel temperature ($T_N \approx 420$ K).[14,19–21] With the influence of spin orbit coupling (SOC), the remarkable structure leads to an anomalous Hall contribution and a vanishingly small magnetization which allows for the realignment of domains by a small magnetic field and large changes of the Hall resistivity at room temperature.[11,14] Similarly, using the theory of Berry-phase curvature, the spin Hall effect is expected to exist in $Mn_3Sn$, where the induced spin currents can be utilized to manipulate the antiferromagnetic magnetic moments or stimulate the motions of domain walls.[15,22] Besides, the feature of topological Weyl semimetals,[18,23] anomalous Nernst effect[24–27] and magneto-optical Kerr effect[28] were found in Mn$_3$Sn single crystals, making Mn$_3$Sn attractive for both scientific research and spintronic application.

Now the research interest is whether the promising magnetotransport properties can be obtained in Mn$_3$Sn films, which is the basis for the device integration and resultant miniaturization. Fortunately, (0001)-oriented Mn$_3$Sn thin films were epitaxially grown on Y:ZrO$_2$ (111) substrates with a Ru underlayer.[29] Nevertheless, the growth mode is composed of approximately 400-nm-wide islands, limiting the transport measurements of the films. As revealed in Mn$_3$Sn single crystals, AHE is strongly dependent on the measurement configuration, both magnetic field (*H*) and current (*I*) with the order of decreasing anomalous Hall resistivity ($\rho_H$): *H* // $[01\bar{1}0]$ & *I* // [0001], *H* // $[2\bar{1}\bar{1}0]$ & *I* // $[01\bar{1}0]$, and *H* // [0001] & *I* // $[01\bar{1}0]$. The first two scenarios show much larger $\rho_H$, because the weak ferromagnetic moment embedded in the chiral antiferromagnetic state is arranged in the plane of (0001), which can be switched by the field in the same plane.[11] Different from the single crystal, one of the challenges is that the current cannot be applied along *c*-axis in (0001)-oriented films for Hall measurements. Also, to avoid the interference of the out-of-plane net magnetic moment on the AHE, we use both $(11\bar{2}0)$- and (0001)-oriented Mn$_3$Sn films to explore AHE, which shows sizable in-plane anomalous Hall conductivity up to room temperature.

Figure 1b shows the sketches of the *L*-shape Hall bar for magnetotransport measurements, where the experimental setup is also included. In this *L*-shape Hall bar, AHE of $(11\bar{2}0)$-oriented films can be measured in two different configurations, i.e., *H* // $[01\bar{1}0]$ & *I* // [0001], and *H* // [0001] & *I* // $[01\bar{1}0]$. Figure 1c presents a typical field dependent Hall resistivity $\rho_H$ of 50 nm-thick $(11\bar{2}0)$-oriented Mn$_3$Sn films, where the experimental conditions are *H* //

$[01\bar{1}0]$ & $I$ // [0001] at 100 K. The $\rho_H$ curve exhibits a clear hysteresis loop with a jump of approximate 1 μΩ cm, which is the same order of magnitude as that of Mn$_3$Sn single crystals.[11] The AHE signals originate from the noncollinear 120° spin structure of antiferromagnetic Mn$_3$Sn, where a very small net magnetic moment rotates along the applied magnetic field.[11] As illustrated in the inset of Figure 1c, the chiral spin configuration of Mn$_3$Sn is reversal when $H$ is swept through zero, giving rise to the sign change of AHE with different $\rho_H$ values.

The situation turns out to be dramatically different for the Hall signal of ferromagnetic films.[30–33] We take a typical ferromagnet TaN/CoFeB/MgO as a control sample. Corresponding Hall curve is displayed in Figure 1d. Since $H$ is applied in plane, the planar Hall resistivity can be expressed as $\rho_H = (\rho_{//} - \rho_{\perp})\sin\varphi\cos\varphi$, where $\rho_{//}$ and $\rho_{\perp}$ are the resistivities for the current to be parallel and perpendicular to the magnetization, respectively, and $\varphi$ is the angle between the magnetization and the current.[33] $\rho_H$ tends to be zero and exhibits a symmetric behavior when a comparatively high magnetic field of $\mu_0 H \approx$ 0.4 T is applied, irrespective of positive or negative field, because of the magnetic moment rotation towards the field direction. Remarkably, the asymmetric AHE of the Mn$_3$Sn films is in strong contrast to the symmetric feature of the Hall effect of ferromagnetic CoFeB, revealing that the observed AHE signals in the Mn$_3$Sn films stem from the chiral spin configuration of antiferromagnets rather than the direct magnetic moment change of ferromagnets.

We now turn towards the AHE of Mn$_3$Sn films with different measurement configurations: $H$ // $[01\bar{1}0]$ & $I$ // [0001], $H$ // [0001] & $I$ // $[01\bar{1}0]$, and $H$ // $[2\bar{1}\bar{1}0]$ & $I$ // $[01\bar{1}0]$. For this experiment, the first two configurations are measured in $(11\bar{2}0)$-oriented films with the $L$-shape Hall bar displayed in Figure 1b, while the last one are recorded in (0001)-oriented films with a normal Hall bar. Concomitant $H$ dependent $\rho_H$ ($\rho_H$–$H$) curves recorded at 200 K are presented in Figure 2a. The most eminent feature is all of the curves show magnetization hysteresis loop-like behavior, instead of the symmetric one in ferromagnetic systems, indicating the switching of the chiral ferromagnetic states by magnetic fields.[11,14] Also visible is the sequence of decreasing $\rho_H$: $H$ // $[01\bar{1}0]$ & $I$ // $[0001]$, $H$ // $[2\bar{1}\bar{1}0]$ & $I$ // $[01\bar{1}0]$, and $H$ // $[0001]$ & $I$ // $[01\bar{1}0]$, coinciding with its counterpart in the Mn$_3$Sn single crystal.[11] Note that unexpected AHE also exists for the $H$ // $[0001]$ case, in contrast to the linear diamagnetic-like $\rho_H$–$H$ curve of single crystal,[11] which will be discussed below.

Corresponding Hall conductivity curves as a function of magnetic field ($\sigma_H$–$H$) at 200 K are displayed in Figure 2b. The Hall conductivity $\sigma_H$ is calculated by $\sigma_H = -\rho_H/\rho^2$, where $\rho$ is the longitudinal resistivity. The $\sigma_H$–$H$ curves also exhibit a clear jump near zero field and hysteresis because the longitudinal resistivity $\rho$ keeps almost constant during the switching process. Interestingly, qualitatively similar but reduced $\sigma_H$–$H$ curves can be measured up to room temperature, as presented in Figure 2c. Note that the curve measured at $H$ // $[2\bar{1}\bar{1}0]$ & $I$ // $[01\bar{1}0]$ shows the highest $\sigma_H$, in contrast to the lowest one for the $H$ // $[0001]$ & $I$ // $[01\bar{1}0]$ case with only half of the $\sigma_H$. Such a difference reaffirms that the AHE signals are correlated to chiral spin order of the Mn$_3$Sn films. A comparison of the $\rho_H$–$H$ curves at three typical temperatures measured in the configuration of $H$ // $[01\bar{1}0]$ & $I$ // $[0001]$ is shown in Figure 2d. As expected, both the magnitude and the hysteresis window of the $\rho_H$–$H$ curves are reduced with increasing temperature, but the jump of the AHE signals remains at 300 K.

To understand the magnetotransport of the Mn$_3$Sn films, their morphology and microstructure are characterized. The surface roughness of the 50 nm-thick $(11\bar{2}0)$- and (0001)-oriented Mn$_3$Sn films, measured by atomic force microscope, is 2.68 and 1.69 nm, respectively, indicating that the whole film is continuous. This feature provides a pre-condition for the AHE measurements as discussed above. Figure 3a shows x-ray diffraction (XRD) spectra of the Mn$_3$Sn films grown on Al$_2$O$_3$ $(1\bar{1}02)$ substrate. In Figure 3a, the diffraction peaks from Mn$_3$Sn $(11\bar{2}0)$ planes can be clearly seen, indicating the quasi-epitaxial growth of $(11\bar{2}0)$-oriented Mn$_3$Sn films. Moreover, there is no secondary phase within the sensitivity of XRD measurements. Although the strong Mn$_3$Sn $(11\bar{2}0)$ texture is detected, it is difficult to imagine the edge-to-edge epitaxial growth for the Al$_2$O$_3$ $(1\bar{1}02)$ due to the large lattice mismatch of 7.4% between them. To understand the growth mode, a comparison of the Φ scan obtained from the $\{20\bar{2}1\}$ planes of the Mn$_3$Sn films and from $\{11\bar{2}0\}$ planes of Al$_2$O$_3$ substrates is displayed in Figure 3b. The crystallographic orientation relationship is determined as Mn$_3$Sn $(11\bar{2}0)$ $[01\bar{1}0]$ // Al$_2$O$_3$ $(1\bar{1}02)$ $[11\bar{2}0]$. An inspection of the Φ scan shows that the peaks are separated by 90° for the film with the fourfold symmetry, which is different from the 180° separation of the Φ scan of the substrate. The transition of twofold symmetry of the substrate to the fourfold symmetry of the film reflects the rotation of crystalline grains for 90° in the film plane, as illustrated in the inset of Figure 3b, which is beneficial for decreasing the interfacial strain and resultant quasi-epitaxial growth, as the mismatch between Mn$_3$Sn $(11\bar{2}0)$[0001] and Al$_2$O$_3$ $(1\bar{1}02)$ $[11\bar{2}0]$ is up to 4.8%. Although the film shows the fourfold symmetry, only a small part of crytalline grains is expected to be rotated 90° in the film plane, taking more robust anomalous Hall conductivity for the case of *H* // $[01\bar{1}0]$ & *I* // [0001] than its *H* // [0001] & *I* // $[01\bar{1}0]$ counterpart into account. Otherwise their AHE would be equal.

A similar characterization was carried out for the Mn$_3$Sn films grown on MgO (111) substrates. Concomitant XRD spectrum is presented in Figure 3c, where the Mn$_3$Sn films exhibits a strong (0001) preferred orientation. According to the Φ scan in Figure 3d acquired from the $\{20\bar{2}1\}$ planes of the Mn$_3$Sn films and from $\{101\}$ planes of MgO substrate, the quasi-epitaxial relationship is Mn$_3$Sn (0001) $[11\bar{2}0]$ // MgO (111) $[1\bar{1}0]$. Also visible is that the peaks are separated 60° for the films, compared to 120° of the substrates. However, this feature does not mean the in-plane rotation of the Mn$_3$Sn (0001) films, instead, it is quite characteristic for the sixfold symmetry of the Mn$_3$Sn (0001) films and the threefold symmetry of the MgO (111) substrate.[29] A typical Mn$_3$Sn/MgO cross-sectional high resolution transmission electron microscopy (HRTEM) image is presented in Figure 3e. The quasi-epitaxial relationship of Mn$_3$Sn (0001) // MgO (111) can be detected from the image. Note that there are fluctuations for the Mn$_3$Sn (0001) planes, producing extensive structural defects in the films, such as dislocations and stacking faults. Low magnification image of the cross-section specimens in Figure 3f unravels the uniform thickness of ~50 nm, ensuring the magnetotransport measurements as discussed above.

The magnetic properties of the Mn$_3$Sn measured by a superconducting quantum interference device (SQUID) magnetometry are shown in Figure 4. Figure 4a shows two magnetization curves of the $(1 1\bar{2} 0)$-oriented Mn$_3$Sn films at 300 K. The magnetic field are applied in two in-plane directions, [0001] and $[0 1\bar{1} 0]$, consistent with measurement configurations for the AHE curves. At a glance, both curves exhibit a diamagnetic behavior attributed to the diamagnetic background of the substrate, indicating the antiferromagnetic feature of the Mn$_3$Sn $(1 1\bar{2} 0)$ films. Corresponding magnetization curves after the subtraction of the diamagnetic background are displayed in Figure 4b. Remarkably, weak but clearly detected hysteresis loops, ~2 and ~3 emu cm$^{-3}$, respectively, can be obtained for both cases of *H* // [0001] and *H* // $[0 1\bar{1} 0]$. The weak magnetization along the $[0 1\bar{1} 0]$ direction can be converted to 0.041 $\mu_B$/per formula unit (f.u.), which is slightly larger than that of single crystal at room temperature.[11,14] The magnetization curve (*H* // $[2 \bar{1} \bar{1} 0]$) of the Mn$_3$Sn (0001) films deposited on the MgO (111) substrate also exhibits a diamagnetic character, due to the diamagnetic background of the MgO (111) substrate. Nevertheless a small kink can be clearly seen in the linearly diamagnetic curve in Figure 4c, indicating a stronger ferromagnetic signal of ~30 emu cm$^{-3}$ in this sample. This can be disclosed by the hysteresis loop shown in Figure 4d after the background subtraction.

The theoretical calculation using Berry curvature and the magnetotransport measurements of Mn$_3$Sn single crystal show that there is no AHE if the magnetic field is applied along [0001] (*c*-axis).[11] In our case, the existence of AHE for *H* // [0001] is most likely correlated to the structural defects in the Mn$_3$Sn films, as revealed by the HRTEM images in Figure 3e. The introduction of the defects is to compensate large interfacial strain, due to the large lattice mismatch between the Mn$_3$Sn films and the substrates as well as the magnetron sputtering technique itself.[29] The existing defects would break the symmetry of some Mn$_3$Sn crystal cells, causing the uncompensated magnetization of Mn atoms, even along [0001], and then generating the undesirable AHE in this direction. Although the unexpected AHE exists along [0001], more robust anomalous Hall conductivity is visible in the other two directions of $[01\bar{1}0]$ and $[2\bar{1}\bar{1}0]$, indicating that the AHE in the present Mn$_3$Sn films is reliable.

In conclusion, robust anomalous Hall signals are observed in quasi-epitaxial Mn$_3$Sn thin films up to room temperature. Three different measurement configurations are used in $(11\bar{2}0)$- and (0001)-oriented Mn$_3$Sn films to confirm the existence of anomalous Hall effect. The anomalous Hall resistivity or conductivity for *H* // $[01\bar{1}0]$ & *I* // [0001] and *H* // $[2\bar{1}\bar{1}0]$ & *I* // $[01\bar{1}0]$ scenarioes is greater than their *H* // [0001] & *I* // $[01\bar{1}0]$ counterpart. The weak ferromagnetic moment in the (0001) plane is responsible for the switching of the Hall signals with magnetization reversal. The integration of Mn$_3$Sn in antiferromagnetic spintronics is strongly expected with higher anomalous Hall conductivity through optimizing the growth of Mn$_3$Sn films.

**Experimental Section**

*Sample preparation*: 50 nm-thick $(11\bar{2}0)$- and (0001)-oriented Mn$_3$Sn films were grown by magnetron sputtering on single crystal Al$_2$O$_3$ $(1\bar{1}02)$ and MgO (111) substrates, respectively. The optimized growth temperature was 420 °C. The base pressure was $2 \times 10^{-5}$ Pa and the process gas (Ar) pressure was 0.4 Pa, with the growth rate being 0.16 nm/s. Considering the different sputtering yields of Mn and Sn elements as well as the fact that Mn$_3$Sn was stable only in the presence of excess Mn,[34,35] a Mn-rich Mn$_3$Sn target was used for the film growth.

*Sample characterization*: X-ray diffraction (XRD) of the Mn$_3$Sn films was measured using Cu *Kα*1 radiation with $\lambda = 1.5406$ Å. High resolution transmission electron microscopy (HRTEM) was used to characterizae the microstructure of film/substrate cross-section. The surface roughness was characterized by atomic force microscope (AFM). Magnetic properties were measured by a superconducting quantum interference device (SQUID) magnetometry at 300 K with the field up to 5 Tesla. *L*-sharpe Hall-bar sketched in Figure 1b and a normal Hall bar were used to conduct the magnetotransport measurements for $(11\bar{2}0)$- and (0001)-oriented Mn$_3$Sn films, respectively. The magnetic field is up to 9 Tesla with the temprature ranging from 100 to 300 K.

**Acknowledgements:**

We acknowledge the support of Beijing Innovation Center for Future Chip (ICFC) and Young Chang Jiang Scholars Program. This work was supported by the National Key R&D Program of China (Grant Nos. 2017YFB0405704) and the National Natural Science Foundation of China (Grant Nos. 51671110, 51571128, and 51871130).

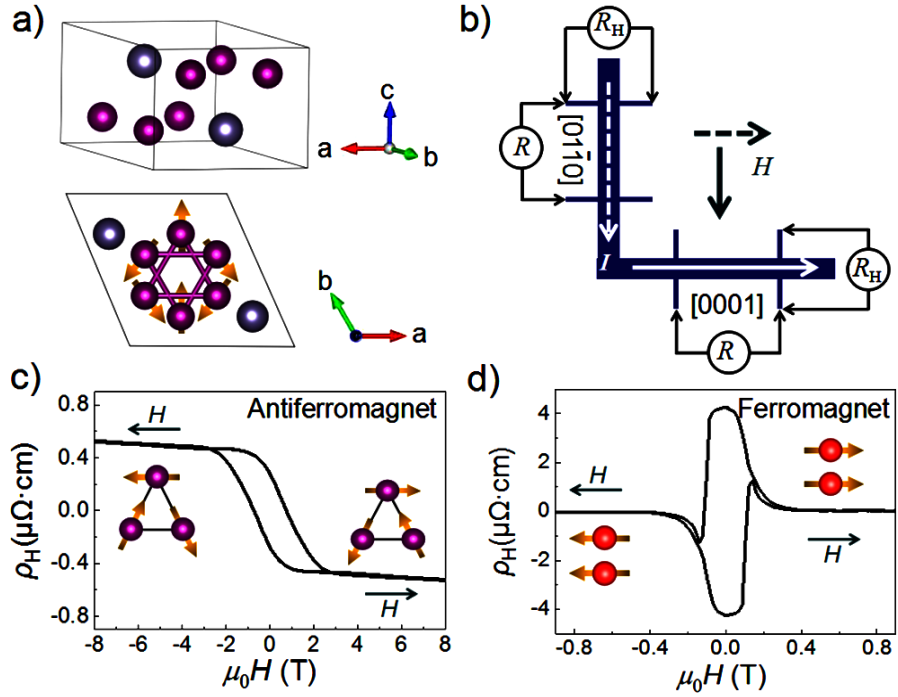

**Figure 1**. a) Crystal and magnetic structure of $Mn_3Sn$, where the purple and grey spheres represent the Mn and Sn atoms, respectively. b) Schematic of the Hall bar devices of $(11\bar{2}0)$-oriented $Mn_3Sn$ films. The solid arrow shows the measurement configuration of $H // [01\bar{1}0]$ & $I // [0001]$. The dashed arrow shows the measurement configuration of $H // [0001]$ & $I // [01\bar{1}0]$. Field dependence of the Hall resistivity $\rho_H$ of c) antiferromagnetic $Mn_3Sn$ film and d) ferromagnetic TaN/CoFeB/MgO/Ta film. The inset illustrates the spin structure of them when $H$ is applied along opposite directions.

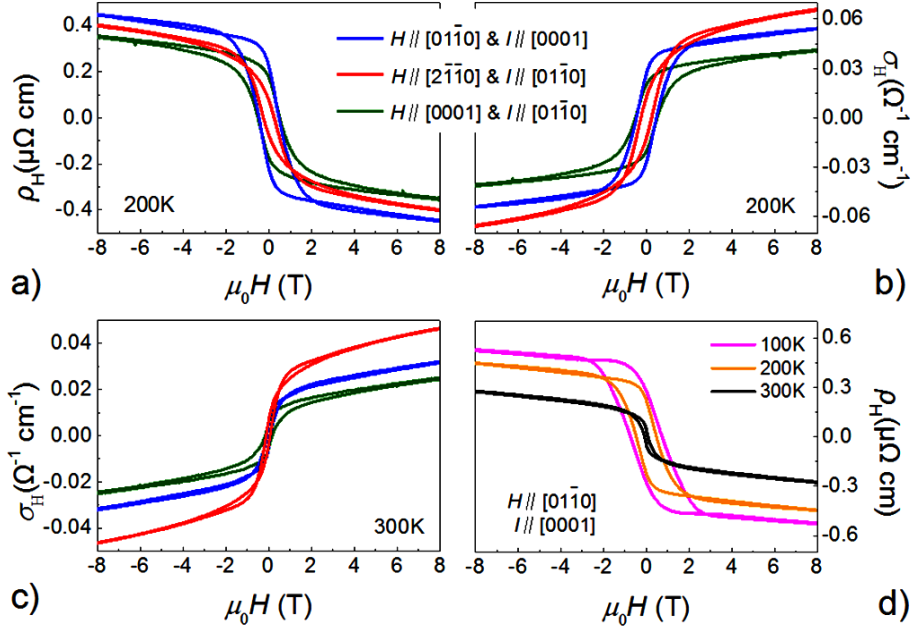

**Figure 2**. a) Magnetization dependence of $\rho_H$ measured with the configuration of $H$ // $[01\bar{1}0]$ & $I$ // $[0001]$, $H$ // $[2\bar{1}\bar{1}0]$ & $I$ // $[01\bar{1}0]$, and $H$ // $[0001]$ & $I$ // $[01\bar{1}0]$ at 200 K. b) and c) The anomalous Hall conductivity $\sigma_H$ versus $H$ obtained at 200 K and 300 K, respectively. d) Field dependence of the Hall resistivity $\rho_H$ at typical temperatures, 100, 200, and 300 K, with the measurement configuration of $H$ // $[01\bar{1}0]$ & $I$ // $[0001]$.

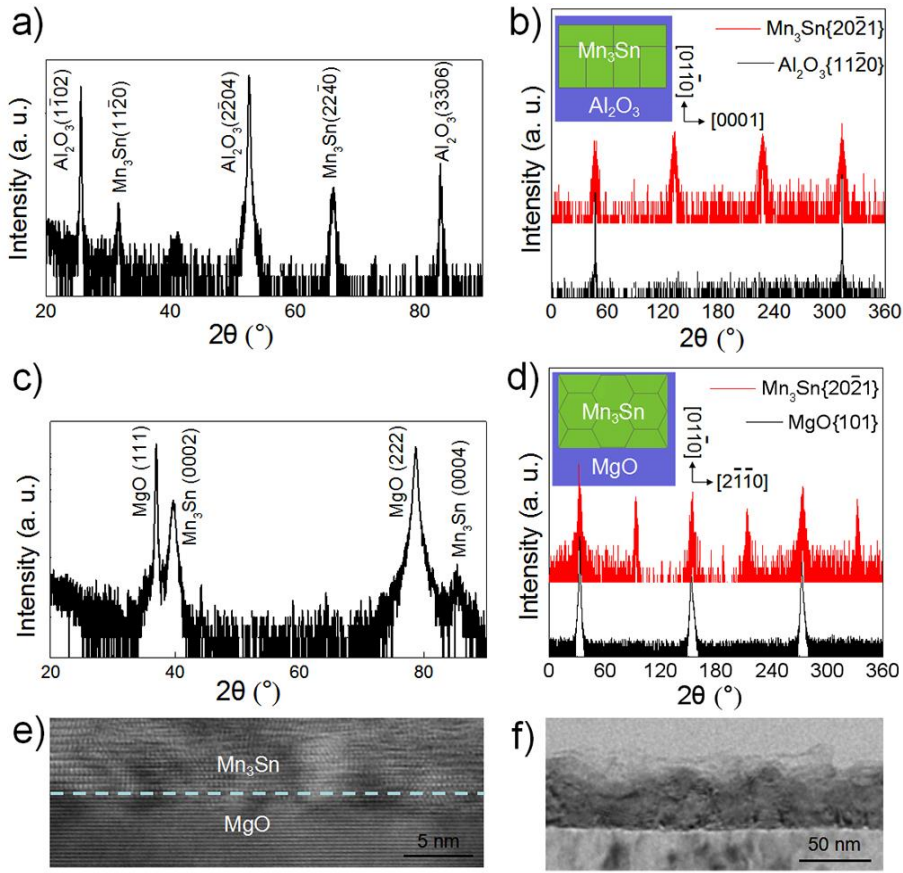

**Figure 3**. XRD patterns of the 50 nm Mn$_3$Sn films on a) Al$_2$O$_3$ ($1\bar{1}02$) substrate and c) MgO (111) substrate. Φ scan patterns of the {$20\bar{2}1$} planes from the Mn$_3$Sn films and b) {$11\bar{2}0$} planes from the Al$_2$O$_3$ substrate and d) {101} planes from the MgO substrate. The insets of b) and d) illustrate the in-plane distribution of the films on different substrates. e) HRTEM image of the interface between the Mn$_3$Sn film and the MgO substrate. f) Low magnification image of the cross-sectional specimens.

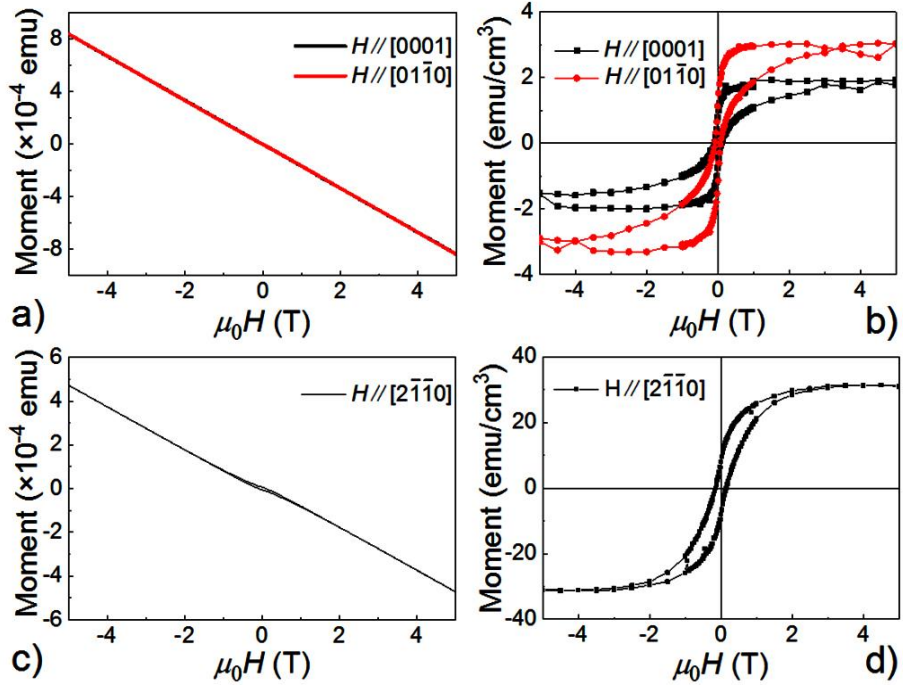

**Figure 4**. Magnetization hysteresis loops of Mn$_3$Sn films deposited on Al$_2$O$_3$ $(1\bar{1}02)$ substrate a) with and b) without diamagnetic background from the substrate for $H \parallel [01\bar{1}0]$ and $H \parallel [0001]$ in the film plane, respectively. Magnetization hysteresis loops of the Mn$_3$Sn films grown on MgO (111) substrate c) with and d) without diamagnetic background from the substrate, where $H \parallel [2\bar{1}\bar{1}0]$ was applied.